\documentclass[twocolumn,showpacs,preprintnumbers,amsmath,amssymb]{revtex4}

\usepackage{graphicx}
\usepackage{dcolumn}
\usepackage{bm}
\usepackage{amsmath}

\begin{document}
\title{Power-law Strength-Degree Correlation From a Resource-Allocation Dynamics on Weighted Networks}
\author{Qing Ou$^{1,2}$}
\author{Ying-Di Jin$^{1}$}
\author{Tao Zhou$^{1}$}
\email{zhutou@ustc.edu}
\author{Bing-Hong Wang$^{1}$}
\author{Bao-Qun Yin$^{2}$}
\affiliation{%
$^{1}$Department of Modern Physics and Nonlinear Science Center\\
$^{2}$Department of Automation, University of Science and Technology of China, 230026, PR China  \\
}%

\date{\today}

\begin{abstract}
Many weighted scale-free networks are known to have a power-law
correlation between strength and degree of nodes, which, however,
has not been well explained. We investigate the dynamic behavior
of resource/traffic flow on scale-free networks. The dynamical
system will evolve into a kinetic equilibrium state, where the
strength, defined by the amount of resource or traffic load, is
correlated with the degree in a power-law form with tunable
exponent. The analytical results agree well with simulations.
\end{abstract}


\maketitle

\section{Introduction}
A very interesting empirical phenomenon in the study of weighted
networks is the power-law correlation between strength $s$ and
degree $k$ of nodes $s \sim k^\theta$
\cite{Li2004,Barrat2004a,Wang2005a,Liu2006}. Very recently, Wang
\emph{et al} have proposed a mutual selection model to explain the
origin of this power-law correlation \cite{Wang2005b}. This model
can provide a partial explanation for social weighted networks,
that is, although the general people want to make friend with
powerful men, these powerful persons may not wish to be friendly
to them. However, this model can not explain the origin of
power-law strength-degree correlation in weighted technological
networks.

In many cases, the concepts of edge-weight and node-strength are
associated with network dynamics. For example, the weight in
communication networks is often defined by the load along with the
edge \cite{Newman2004}, and the strength in epidemic contact
networks is defined by the individual infectivity \cite{Zhou2006}.
On the one hand, although the weight/strength distribution may
evolve into a stable form, the individual value is being changed
with time by the dynamical process upon network. On the other
hand, the weight/strength distribution will greatly affect the
corresponding dynamic behaviors, such as the epidemic spreading
and synchronization \cite{Yan2005,Motter2005,Chavez2005,Zhao2006}.

Inspired by the interplay of weight and network dynamics, Barrat,
Barth\'elemy, and Vespignani proposed an evolution model (BBV
model for short) for weighted networks
\cite{Barrat2004b,Barrat2004c}. Although this model can naturally
reproduce the power-law distribution of degree, edge-weight, and
node-strength, it fails to obtain the power-law correlation
between strength and degree. In BBV model, the dynamics of weight
and network structure are assumed in the same time scale, that is,
in each time step, the weight distribution and network topology
change simultaneously. Here we argue that the above two time
scales are far different. Actually, in many real-life situations,
the individual weight varies momently whereas the network topology
only slightly changes during a relatively long period. Similar to
the traffic dynamics based on the local routing protocol
\cite{Holme2003,Tadic2004,Yin2006,Wang2006}, we investigate the
dynamic behaviors of resource/traffic flow on scale-free networks
with given structures, which may give some illuminations about the
origin of power-law correlation between strength and degree in
weighted scale-free networks.

\section{Resource flow with preferential allocation}
As mentioned above, strength usually represents resources or
substances allocated to each node, such as wealth of individuals
of financial contact networks \cite{Xie2005}, the number of
passengers in airports of world-wide airport networks
\cite{Guimera2005}, the throughput of power stations of electric
power grids \cite{Albert2004}, and so on. These resources also
flow constantly in networks: Money shifts from one person to
another by currency, electric power is transmitted to every city
from power plants by several power hubs, and passengers travel
from one airport to another. Further more, resources prefers to
flow to larger-degree nodes. In transport networks, large nodes
imply hubs or centers in traffic system. So passengers can get a
quick arrival to the destinations by choosing larger airports or
stations. In financial systems, people also like to buy stocks of
larger companies or deposit more capital in the banks with more
capital because larger companies and banks generally have more
power to make profits and more capacity to avoid losses. Inspired
by the above facts, we propose a simple mechanism to describe the
resource flow with preferential allocation in networks.

\begin{figure}
\scalebox{0.5}[0.5]{\includegraphics{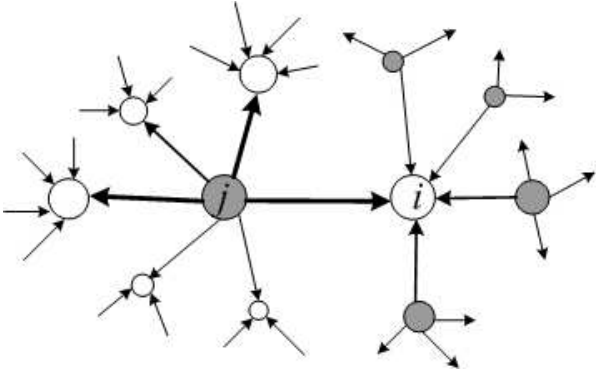}}
\caption{\label{fig:resource flow} Resources in node $j$ are
divided into several pieces and then flow to its neighbors. The
thicker lines imply there are more resources flowing. It is worth
pointing out that, in order to give a clearer illustration we do
not plot the resource flow into node $j$ or out of node $i$.}
\end{figure}

At each time, as shown in Fig. 1, resources in each node are
divided into several pieces and then flow to its neighbors. The
amount of each piece is determined by its neighbors' degrees. We
can regulate the extent of preference by a tunable parameter
$\alpha$. The equations of resource flow are
\begin{equation}
Q_{j\rightarrow i}(t)= {k^{\alpha}_{i}s_{j}(t)}/{\sum\limits_{l\in
N(j)}k^{\alpha}_l},
\end{equation}
where $Q_{j \rightarrow i}(t)$ is the amount of resources moving
from node $j$ to $i$ at time $t$, $s_{j}(t)$ is the amount of
resources owned by node $j$ at time $t$, $k_i$ is the degree of
node $i$ and $N(j)$ is the set of neighbors of node $j$. If $i$
and $j$ are not neighboring, then $Q_{j\rightarrow
i}=Q_{i\rightarrow j}=0$. Meanwhile each node also gets resources
from its neighbors, so at time $t+1$ ,$\forall$ $i$
\begin{equation}\label{strength of node i}
\begin{split}
s_{i}(t+1)=\sum_{j \in N(i)}Q_{j \rightarrow i}(t) &=\sum_{j \in
N(i)} \big({k^{\alpha}_{i}s_{j}(t)}/{\sum_{l\in
N(j)}k^{\alpha}_l}\big).
\end{split}
\end{equation}

\begin{figure}
\scalebox{0.8}[1]{\includegraphics{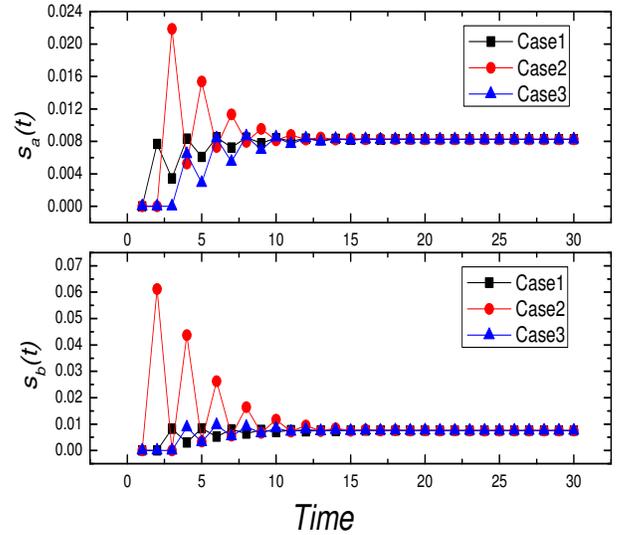}}
\caption{\label{fig:converge} (Color online) The evolution of the
strengths of node $a$ and $b$, where nodes $a$ and $b$ are
randomly selected for observation. The three cases are in
different initial states which simply satisfy $\sum_i{s_i(0)=1}$.
The exponent $\alpha=1$.}
\end{figure}
\begin{figure}
\scalebox{0.8}[0.8]{\includegraphics{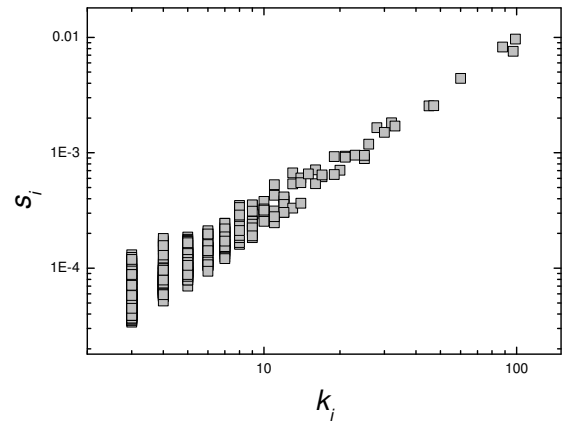}}
\caption{\label{fig:SiKi} Scatter plots of $s_i$ vs $k_i$ for all
the nodes}
\end{figure}

\section{Kinetic equilibrium state}
The Eq. \ref{strength of node i} can be expressed in terms of a
matrix equation, which reads
\begin{equation}\label{matrix}
\vec{S}(t+1)=A\vec{S}(t):=\begin{pmatrix} a_{11} & \dots &
a_{1n}\\
a_{21} & \dots & a_{2n}\\
\hdotsfor{3}\\
a_{n1} & \dots & a_{nn}\\
\end{pmatrix}
\cdot
\begin{pmatrix} s_{1}(t)\\
s_{2}(t)\\
\dots\\
s_{n}(t)
\end{pmatrix}
\end{equation}
where the elements of matrix $A$ are given by
\begin{equation}
a_{ij}=
\begin{cases}
{k^{\alpha}_i}/{\sum\limits_{l \in N(j)}k^{\alpha}_l}& j \in N(i)\\
0& \text{otherwise}
\end{cases}
\end{equation}
Since $\sum\limits_{i=1}^n |a_{ij}|=1$, $\forall j$,  the spectral
radius of matrix $A$ obeys the equality $\rho(A)\leq 1$, according
to the Gershg\"{o}rin disk theorem \cite{disk theorem}. Here, the
spectral radius, $\rho(A)$, of a matrix $A$, is the largest
absolute value of an eigenvalue. Further more, since the
considered network is symmetry-free (That is to say, the network
is strongly connected thus for any two nodes $i$ and $j$, these
exists at least one path from $i$ to $j$), $A^k$ will converge to
a constant matrix for infinite $k$. That is, if given the initial
boundary condition to Eq. \ref{matrix} (e.g. let
$\sum\limits_{i=1}^{n}{s_i(0)=1}$, where $n$ denotes the total
number of nodes in network), then $s_i(t)$ will converge in the
limit of infinite $t$ as $\lim\limits_{t \rightarrow
\infty}s_i(t)=s_i$ for each node $i$.

Consequently, Denote $\vec{S}:=(\begin{matrix} s_1,& s_2 & \dots &
s_n \end{matrix})^T$, one can obtain
\begin{equation}\label{kinetic equilibrium}
\vec{S}=A\vec{S}.
\end{equation}
That is, for any $i$,
\begin{equation}\label{node equilibrium}
s_{i}=\sum\limits_{j \in N(i)}\big({k^{\alpha}_{i}
s_{j}}/{\sum\limits_{l\in N(j)}k^{\alpha}_l}\big).
\end{equation}
From Eq. \ref{kinetic equilibrium}, it is clear that $\vec{S}$ is
just the kinetic equilibrium state of the resource flow in our
model. Since $\vec{S}=\lim\limits_{k \rightarrow
\infty}A^{k}\vec{S}(0)$, $\vec{S}$ is determined only by matrix
$A$, if given the initial boundary condition with $\vec{S}(0)$
satisfying $\sum\limits_{i=0}^{n}{s_i(0)= 1}$. Since matrix $A$ is
determined by the topology only, for each node $i$ in the kinetic
equilibrium, $s_i=\lim\limits_{t \rightarrow \infty}s_i(t)$ is
completely determined by the network structure. $s_i$ denotes the
amount of resource eventually allocated to node $i$, thus it is
reasonable to define $s_i$ as the strength of node $i$.

\begin{figure}
\scalebox{0.8}[0.8]{\includegraphics{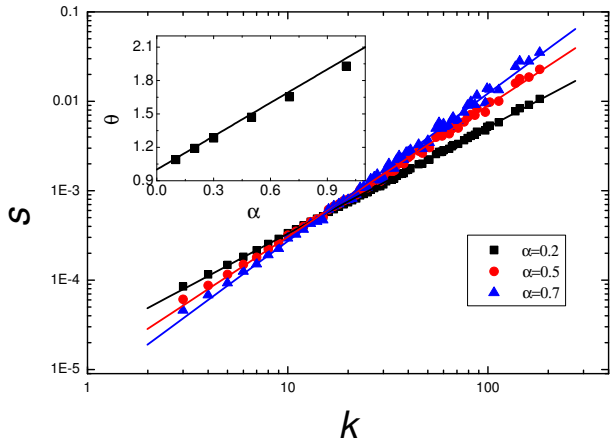}}
\caption{\label{fig:SvsK} (Color online) The correlation between
degree and strength with different $\alpha$. In the inset, the
relation between $\theta$ and $\alpha$ is given, where the squares
come from the simulations and the solid line represents the
theoretical result $\theta=1+\alpha$.}
\end{figure}

\begin{figure}
\scalebox{0.8}[0.8]{\includegraphics{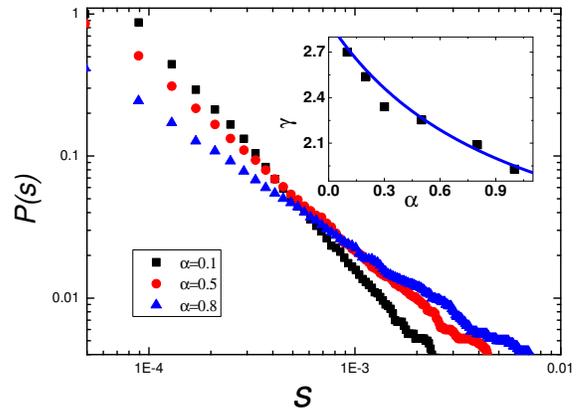}}
\caption{\label{fig:strength} (Color online) The distribution of
strength with different $\alpha$. The inset exhibits the relation
between $\gamma$ and $\alpha$, where the squares come from the
simulations and the solid line represents the theoretic analysis
$\gamma=(\alpha+\beta)/(1+\alpha)$.}
\end{figure}

\section{Power-law correlation between strength and degree in scale-free networks}
The solution of Eq. (6) reads
\begin{equation}\label{strength solution}
s_i=\lambda k^{\alpha}_i \sum\limits_{j \in N(i)}{k^{\alpha}_j},
\end{equation}
where $\lambda$ is a normalized factor.

In principle, this solution gives the analytical relation between
$s_i$ and $k_i$ when $\sum_{j \in N(i)}{k^{\alpha}_j}$ can be
analytically obtained from the degree distribution. For
uncorrelated networks \cite{Newman2002}, statistically speaking we
have
\begin{equation}
s_i=\lambda k^{1+\alpha}_i \sum_{k'}P(k')k'^{\alpha},
\end{equation}
where $P(k)$ denotes the probability a randomly selected node is
of degree $k$. Since $\lambda \sum_{k'}P(k')k'^{\alpha}$ is a
constant when given a network structure, one has $s_i\sim
k_i^{1+\alpha}$, thus
\begin{equation}
s(k)\sim k^{1+\alpha},
\end{equation}
where $s(k)$ denotes the average strength over all the nodes with
degree $k$.

This power-law correlation $s(k) \sim k^{\theta}$ where
$\theta=1+\alpha$, observed in many real weighted networks, can be
considered as a result of the conjunct effect of the above
power-law correlation and the scale-free property. Obviously, if
the degree distribution in a weighted network obeys the form
$P(k)\sim k^{-\beta}$, one can immediately obtain the distribution
of the strength
\begin{equation}
P(s)\sim s^{-\gamma},
\end{equation}
where the power-law exponent $\gamma=(\alpha+\beta)/(1+\alpha)$.

\section{Simulations}
Recent empirical studies in network science show that many
real-life networks display the scale-free property \cite{Review},
thus we use scale-free networks as the samples. Since the
Barab\'asi-Albert (BA) model \cite{BA} is the mostly studied model
and lacks structural-biases such as non-zero degree-degree
correlation, we use BA network with size $n=5000$ and average
degree $\langle k\rangle=6$ for simulations. The dynamics start
from a completely random distribution of resource. As is shown in
Fig. 2, we randomly pick two nodes $a$ and $b$, and record their
strengths vs time $s_a(t)$ and $s_b(t)$ for three different
initial conditions. Clearly, the resource owned by each node will
reach a stable state quickly. And no matter how and where the one
unit resource flow in, the final state is the same.

\begin{figure}
\scalebox{0.8}[0.8]{\includegraphics{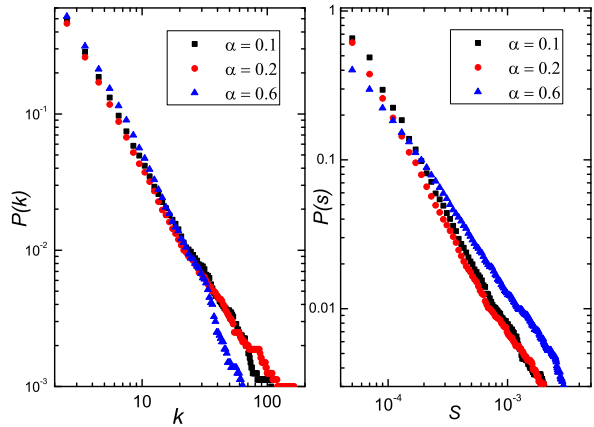}}
\caption{\label{fig:strength} (Color online) The distributions of
degree (left panel) and strength (right panel) with different
$\alpha$. The networks are generated by the strength-PA mechanism,
and those shown here are the sampling of size $n=5000$.}
\end{figure}

\begin{figure}
\scalebox{0.8}[0.8]{\includegraphics{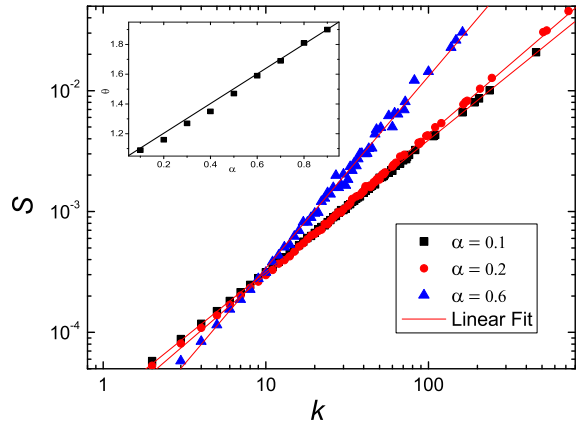}}
\caption{\label{fig:strength}  (Color online) The correlation
between degree and strength with different $\alpha$. In the inset,
the relation between $\theta$ and $\alpha$ is given, where the
squares come from the simulations and the solid line represents
the theoretical result $\theta=1+\alpha$. The networks are
generated by the strength-PA mechanism, and those shown here are
the sampling of size $n=5000$.}
\end{figure}

Similar to the mechanism used to judge the weight of web by
Google-searching (see a recent review paper \cite{Google} about
the \emph{PageRank Algorithm} proposed by Google), the strength of
a node is not only determined by its degree, but also by the
strengths of its neighbors (see Eq. 7). Although statistically
$s(k)\sim k^{1+\alpha}$ for uncorrelated networks, the strengths
of the nodes with the same degree may be far different especially
for low-degree nodes as exhibited in Fig. 3.

In succession, we average the strengths of nodes with the same
degree and plot Fig. 4 to verify our theoretical analysis that
there is a power-law correlation $s\sim k^{\theta}$ between
strength and degree, with exponent $\theta=1+\alpha$. Fig. 5 shows
that the strength also obeys power-law distribution, as observed
in many real-life scale-free weighted networks. And the
simulations agree well with analytical results.

\section{Conclusion remarks}
In this paper, we proposed a model for resource-allocation
dynamics on scale-free networks, in which the system can approach
to a kinetic equilibrium state with power-law strength-degree
correlation. If the resource flow is unbiased (i.e. $\alpha=0$),
similar to the BBV model \cite{Barrat2004b,Barrat2004c}, the
strength will be linearly correlated with degree as $s(k)\sim k$.
Therefore, the present model suggests that the power-law
correlation between degree and strength arises from the mechanism
that resources in networks tend to flow to larger nodes rather
than smaller ones. This preferential flow has been observed in
some real traffic systems. For example, very recently, we
investigated the empirical data of Chinese city-airport network,
where each node denote a city, and the edge-weight is defined as
the number of passengers travelling along this edge per week
\cite{Liu2006}. We found that the passenger number from one city
to its larger-degree neighbor is much larger than that from this
city to its smaller-degree neighbor. In addition, in Chinese
city-airport network \cite{Liu2006} and US airport network
\cite{Barrat2004a}, the power-law exponents are $\theta\approx
1.4$ and $\theta\approx 1.5$, respectively, which is within the
range of $\theta$ predicted by the present model.

The readers should be warned that the analytical solution shown in
this paper is only valid for static networks without any
degree-degree correlation. However, we have done some further
simulations about the cases of growing networks (see Appendix A)
and correlated networks (see Appendix B). The results are
quantitatively the same with slight difference in quantity.

\begin{figure}
\scalebox{0.9}[0.9]{\includegraphics{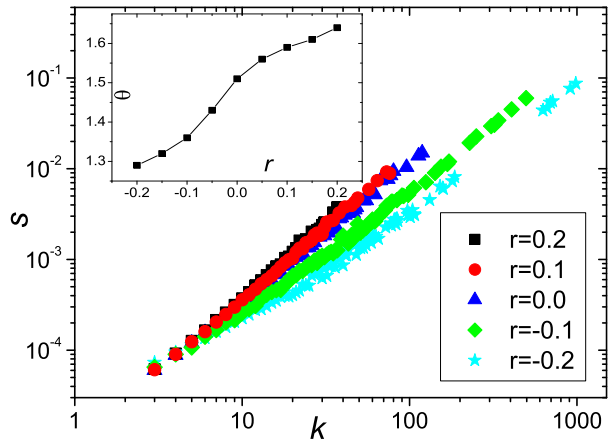}}
\caption{\label{fig:strength}  (Color online) The correlation
between degree and strength with different assortative
coefficients $r$. The parameter $\alpha=0.5$ is fixed. The inset
shows the numerically fitting value of $\theta$ vs assortative
coefficients. The networks are generated by the generalized BA
algorithm \cite{GBA1,GBA2} of size $n=5000$ and average degree
$\langle k\rangle=6$.}
\end{figure}

Finally, in this model, the resource flow will approach to a
kinetic equilibrium, which is determined only by the topology of
the networks, so we can predict the weight of a network just from
its topology by the equilibrium state. Therefore, our proposed
mechanism can well apply to estimate the behaviors in many
networks. When given topology of a traffic network, people can
easily predict the traffic load in individual nodes and links by
using this model, so that this model may be helpful to a better
design of traffic networks.

\begin{acknowledgments}
The authors wish to thank Miss. Ming Zhao for writing the C++
programme that can generate the scale-free networks with tunable
assortative coefficients. This work has been partially supported
by the National Natural Science Foundation of China under Grant
Nos. 70471033, 10472116, and 10635040, the Special Research Founds
for Theoretical Physics Frontier Problems under Grant No.
A0524701, and Specialized Program under President Funding of
Chinese Academy of Science.
\end{acknowledgments}

\appendix

\section{The case of growing networks}
Since many real networks, such as WWW and Internet, are growing
momently. The performance of the present resource-allocation flow
on growing networks is thus of interest. We have implemented the
present dynamical model on the growing scale-free networks
following the usual preferential attachment (PA) scheme of
Barab\'asi-Albert \cite{BA}. Since the topological change is
independent of the dynamics taking place on it, and the relaxation
time before converging to a kinetic equilibrium state is very
short (see Fig. 2), if the network size is large enough (like in
this paper $n\sim 10^3$), then the continued growth of network has
only very slight effect on topology and the results is almost the
same as those of the ungrowing case shown above.

Furthermore, we investigate the possible interplay between the
growing mechanism and the resource-allocation dynamics. In this
case, the initial network is a few fully connected connected
nodes, and the resource is distributed to each node randomly.
Then, the present resource-allocation process works following Eq.
(2), and simultaneously, the network itself grows following a
strength-PA mechanism instead of the degree-PA mechanism proposed
by BA model. That is to say, at each time step, one node is added
into the network with $m$ edges attaching to the existing nodes
with probability proportional to their strengths (In a growing BA
network, the corresponding probability is proportional to their
degrees). Clearly, under this scenarios, there exists strong
interplay between network topology and dynamic.

When the network becomes sufficient large ($n\sim 10^3$), as shown
in Fig. 6, the evolution approaches a stable process with both the
degree distribution and strength distribution approximately
following the power-law forms. Furthermore, we report the
relationship between strength and degree in Fig. 7, which indicate
that the power-law scaling, $s(k)\sim k^{\theta}$ with
$\theta=1+\alpha$, also holds even for the growing networks with
strong interplay with the resource-allocation dynamics.

\section{The case of correlated networks}
Note that, the Eq. (8) is valid under the assumption that the
underlying network is uncorrelated. However, many real-life
networks exhibit degree-degree correlation in some extent. In this
section, we will investigate the case of correlated networks. The
model used in this section is a generalized BA model
\cite{GBA1,GBA2}: Starting from $m_0$ fully connected nodes, then,
at each time step, a new node is added to the network and $m$
($<m_0$) previously existing nodes are chosen to be connected to
it with probability
\begin{equation}
p_i \propto \frac{k_i+k_0}{\sum_j(k_j+k_0)},
\end{equation}
where $p_i$ and $k_i$ denote the choosing probability and degree
of node $i$, respectively. By varying the free parameter $k_0$
$(>-m)$, one can obtain the scale-free networks with different
assortative coefficients $r$ (see the Ref. \cite{Newman2002} for
the definition of assortative coefficient).

The simulation results are shown in Fig. 8, from which one can
find that the power-law correlations between strength and degree
in the correlated networks are quantitatively the same as that of
the uncorrelated networks, however, the power-law exponents,
$\theta$, are slightly different. Actually, in the positive
correlated networks, the large-degree nodes prefer to connect with
some other large-degree nodes rather than those small-degree
nodes, thus there may exist a cluster consisting of large-degree
nodes that can hold the majority of resource. That cluster makes
the large-degree nodes having even more resource than in the
uncorrelated case, thus leading to a larger $\theta$. In the inset
of Fig. 8, one can find that $\theta$ is larger in the positive
correlated networks, and smaller in the negative correlated
networks. However, the analytical solution have not yet achieved
when taking into account the degree-degree correlation, which
needs a in-depth analysis in the future.

\end{document}